\documentclass[
superscriptaddress,
 amsmath,amssymb,
 aps,
showpacs,
prl,
preprint,
]{revtex4}

\usepackage{graphicx}% Include figure files
\usepackage{graphics}
\usepackage{dcolumn}% Align table columns on decimal point
\usepackage{bm}% bold math
\usepackage{hyperref}% add hypertext capabilities
\usepackage{color,hyperref}
\definecolor{blue}{rgb}{0.0,0.0,0.7}
\hypersetup{colorlinks,breaklinks,
           linkcolor=blue,urlcolor=blue,
           anchorcolor=blue,citecolor=blue}

\begin{document}

\title{Evidence for Multiple Chiral Doublet Bands in $^{133}$Ce} 

\author{A.D. Ayangeakaa}
\affiliation{Department of Physics, University of Notre Dame, Notre Dame, IN 46556}%
\author{U. Garg}
\affiliation{Department of Physics, University of Notre Dame, Notre Dame, IN 46556}%
%\affiliation{State Key Laboratory of Nuclear Physics and Technology,
%School of Physics, Peking University, Beijing 100871, China}%
\author{M.D. Anthony}
\author{S. Frauendorf}
\author{J.T. Matta}
\author{B.K. Nayak}
\altaffiliation[Present address: ]{Nuclear Physics Division, Bhabha Atomic Research Centre (BARC), Mumbai 400085, India}
\author{D. Patel}
\affiliation{Department of Physics, University of Notre Dame, Notre Dame, IN 46556}%

\author{Q.B. Chen}
\author{S.Q. Zhang}
\author{P.W. Zhao}
\affiliation{State Key Laboratory of Nuclear Physics and Technology,
School of Physics,
Peking University, Beijing 100871, China}%
\author{B. Qi}
\affiliation{School of Space Science and Physics, Shandong University at Weihai, Weihai 264209, China}
\author{J. Meng}
\affiliation{State Key Laboratory of Nuclear Physics and Technology,
School of Physics,
Peking University, Beijing 100871, China}%
\affiliation{School of Physics and Nuclear Energy Engineering,
Beihang University, Beijing 100191, China}%
\affiliation{Department of Physics,
University of Stellenbosch, Stellenbosch, South Africa}%

\author{R.V.F. Janssens}
\author{M.P. Carpenter}
\affiliation{Physics Division, Argonne National Laboratory, Argonne, IL 60439}
\author{C.J. Chiara}
\affiliation{Physics Division, Argonne National Laboratory, Argonne, IL 60439}
\affiliation{Department of Chemistry and Biochemistry, University of Maryland, College Park, MD 20742}
\author{F.G. Kondev}
\affiliation{Nuclear Engineering Division, Argonne National Laboratory, Argonne, IL 60439}%
\author{T. Lauritsen}
\author{D. Seweryniak}
\author{S. Zhu}
\affiliation{Physics Division, Argonne National Laboratory, Argonne, IL 60439}%
\author{S.S. Ghugre}
\affiliation{UGC-DAE Consortium for Science Research, Kolkata 700 098, India}
\author{R. Palit}
\affiliation{Tata Institute of Fundamental Research, Mumbai 400 005, India}
\affiliation{The Joint Institute for Nuclear Astrophysics, University of Notre Dame, Notre Dame, IN 46556}

\date{\today}

\begin{abstract}
Two distinct sets of chiral-partner bands have been identified in the nucleus $^{133}$Ce. They constitute a multiple chiral doublet (M$\chi$D), a phenomenon predicted by relativistic mean field (RMF) calculations and observed experimentally here for the first time. The properties of these chiral bands are in good agreement with results of calculations based on a combination of the constrained triaxial RMF theory and the particle-rotor model.
\end{abstract}

\maketitle

Described in terms of a coupling between deformation and orientation degrees of freedom,
chirality represents a novel feature of triaxial nuclei rotating
about  an axis that lies outside the three planes spanned by the
principal axes of its mean-field density distribution
~\cite{Frauendorf1997131,dimitrovPRL.84.5732,frauendorf.73.463,meng1,meng2}. For a triaxial nucleus, the short, intermediate, and long principal axes
form a screw with respect to the angular momentum vector,
resulting in the formation of two chiral systems, with left- and
right-handed orientations.
The restoration of the broken chiral symmetry in the laboratory frame
manifests as a pair of degenerate $\Delta I$ = 1 bands with the same parity.
Experimental evidence for such chiral band-pairs has been found in
the $A\sim 190$, $A\sim 130$, $A\sim100$, and $A\sim 80$ mass regions of the
nuclear chart~\cite{Petrache1996106,Starosta.86.971,PhysRevC.63.061304,PhysRevC.63.051302,
PhysRevC.64.031304,PhysRevLett.91.132501,PhysRevC.67.044319,PhysRevLett.92.032501,
Joshi2004135,PhysRevC.69.024317,Timar2004178,PhysRevLett.98.102501,S.Y.Wang2011PLB,chiral190.1,chiral190.2}.

Theoretically, chiral doublet bands have been successfully described  by several formulations {\em viz.} the Triaxial Particle Rotor model (TPRM)~\cite{Frauendorf1997131,Peng2003PRC,
S.Q.Zhang2007PRC,B.Qi2009PLB}; the Tilted Axis Cranking (TAC) model
with shell correction (SCTAC)~\cite{dimitrovPRL.84.5732,Timar2004178,PhysRevC.69.024317}, and Skyrme-Hartree-Fock~\cite{OlbratowskiPRL.93.052501, Olbratowski2006PRC} approaches; and
the random phase approximation (RPA) \cite{MukhopadhyayPRL.99.172501,Almehed2011PRC}.
The general conditions for rotational chirality~\cite{Frauendorf1997131} 
imply that this phenomenon may appear for more than one configuration in the same nucleus. 
The resultant possibility of having multiple pairs of chiral doublet bands in a single nucleus was 
demonstrated for the Rh isotopes by the relativistic mean field (RMF) theory in Refs.~\cite{mengPRC.73.037303,pengPRC.77.024309,J.Li2011PRC},
which introduced the acronym M$\chi$D for multiple chiral doublet bands. The likelihood of chiral bands with different configurations was also discussed in context of SCTAC calculations used to interpret the observed band structures in  $^{105}$Rh~\cite{PhysRevC.69.024317,Timar2004178}.
In this Letter, the first firm experimental evidence for the existence of M$\chi$D is reported in the nucleus $^{133}$Ce. The observations 
represent an important confirmation of triaxial shape coexistence and its
geometrical interpretation.

\begin{figure*}
  \includegraphics[angle=90, scale=1.0]{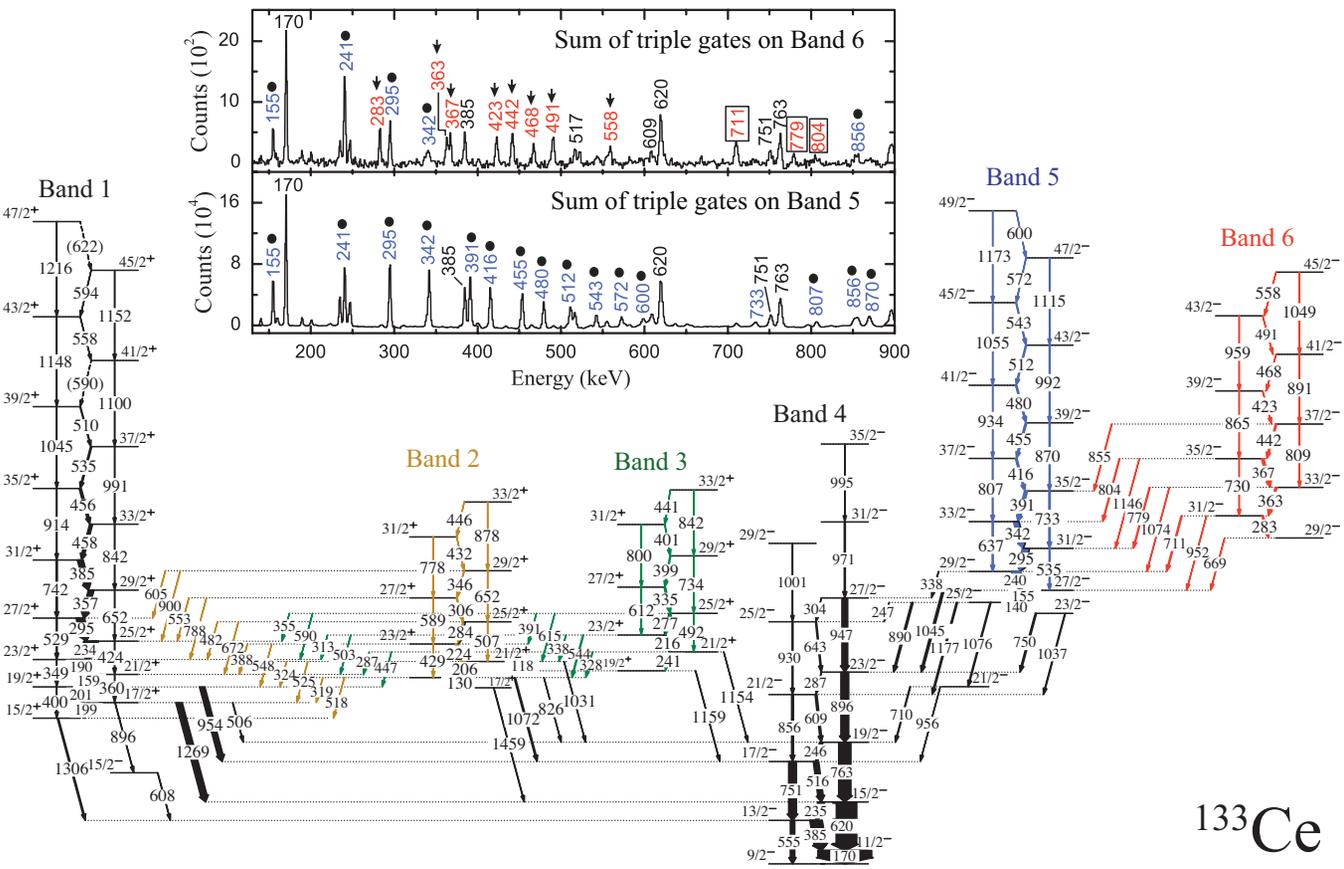}
  \caption{\label{fig:partialscheme}(Color online) Partial level scheme of  $^{133}$Ce
        showing the previously-known bands and the partner bands identified in this work. All transitions within and out of a band are shown in the same color. The lowest level shown ($I^{\pi}=\frac{9}{2}^{-}$) is not the ground state, but an isomer with an excitation energy of 37.2 keV~\cite{nndc}.
        Inset: Background-subtracted $\gamma$-$\gamma$-$\gamma$-$\gamma$ coincidence spectra from  different gate combinations between Bands 5 and 6. Black-, blue-, and red (with arrows)-colored transitions represent members of Bands 4, 5, and 6, respectively. The interconnecting transitions between Bands 5 and 6 are identified by boxes around the transition energies.
         }
\end{figure*}

Two separate experiments were performed using the ATLAS facility at
the Argonne National Laboratory. In both, high-spin states in
$^{133}$Ce were populated following the $^{116}$Cd($^{22}$Ne,5n)$^{133}$Ce
reaction at a bombarding energy of 112 MeV.
In the first experiment, the target was a foil of
isotopically enriched $^{116}$Cd, 1.48 mg/cm$^{2}$-thick, and sandwiched between a 50 $ \mathrm{\mu g/cm^{2}}$-thick
front layer of Al and a 150 $ \mathrm{\mu g/cm^{2}}$-thick Au backing. The second experiment used a target of the same enrichment and thickness
evaporated onto a 55 $ \mathrm{\mu g/cm^{2}}$-thick Au foil. In both experiments,
four-fold and higher prompt $\gamma$-ray coincidence events were measured using
the Gammasphere array~\cite{I.Y.Lee1990NPA}.
A combined total of approximately $4.1\times10^{9}$ four- and higher-fold coincidence events
were accumulated during the two experiments. These events were
analyzed using the RADWARE suite of analysis packages~\cite{Radford1995}.
The level scheme for $^{133}$Ce deduced in the present work builds
substantially upon that previously reported for this nucleus~\cite{Ma-PhysRevC.36.2322,
 Hauschild1995438,Hauschild.54.613}. A partial level scheme comprising the structures relevant to the focus of this Letter is
displayed in Fig.~\ref{fig:partialscheme}. Spin and parity assignments
for newly-identified levels were made on the basis of extensive measurements of DCO ratios~\cite{KramerFlecken1989333,Chiara.75.054305} and,
where practical, confirmed by angular-distribution analyses~\cite{Iacob199757}.

Two strongly coupled $\Delta I =1$, positive-parity bands were established: the lower-energy
band (Band 2 in Fig.~\ref{fig:partialscheme}) and its partner (Band 3) which feeds into the former by a number of strong $\Delta I =1$, and much weaker $\Delta I =2$, ``linking'' transitions. 
Both bands were known previously, but had no
conclusive multipolarity 
and configuration assignments. They have been extended
up to $I^{\pi}=\frac{33}{2}^{+}$ and new in-band crossover
transitions have been added, together with the aforementioned linking transitions.
From DCO ratios and angular-distribution analyses,
the $\Delta I=1$ transitions in Bands 2 and 3 were found to be of the
dipole/quadrupole type; they are assumed to be of $M1/E2$ mixed multipolarity.
The corresponding crossover transitions within the bands were found to
have $E2$ character. Likewise, the multipolarities of the $\Delta I =1$, 328-, 338-,
and 391-keV linking transitions were determined to be of the mixed $M1/E2$ type~\cite{daniel.thesis}. 
This, in conjunction with the deduced $E2$ multipolarity of the
544- and 615-keV transitions, establishes that Bands 2 and 3 have the same parity.
The observation of linking transitions between Bands 2 and 3, and the hitherto-unknown 
multiple decays of the two bands into Band 1 via a series of $\Delta I =1$
and $\Delta I =2$ transitions, suggests an underlying similarity of the quasiparticle
configurations associated with these bands. These decay sequences are interpreted as possibly resulting from
the interaction between Bands 1 and 2 (and, by similar argument, Bands 1
and 3), following detailed calculations based on 
triaxial RMF theory in which the configuration of Band 1 was revised
to $\pi(1h_{11/2})(2d_{5/2})\otimes \nu(h_{11/2}){}^{-1}$ from its previously
assigned $\pi(1g_{7/2})^{-1}(1h_{11/2})\otimes \nu(h_{11/2}){}^{-1}$
configuration~\cite{Ma-PhysRevC.36.2322}.
Consequently, Bands 2 and 3 are assigned the $\pi(1g_{7/2}){}^{-1}(1h_{11/2})\otimes
\nu(1h_{11/2}){}^{-1}$ configuration. 

A second pair of composite bands, consisting of the sequences identified as
Bands 5 and 6 in Fig.~\ref{fig:partialscheme}, was also established, with both bands
again comprising strong $\Delta I=1$ and relatively weak $\Delta I=2$ transitions.
The inset in Fig.~\ref{fig:partialscheme} displays summed spectra obtained using various gate
combinations to highlight the in-band and interconnecting transitions in Bands 5 and 6.
The spins and parity of levels in Band 5, which has been slightly extended by
the addition of new transitions above the 45/2$^{-}$ level, were suggested previously by
Ma et al.~\cite{Ma-PhysRevC.36.2322};
these assignments have been confirmed by the present angular-distribution and DCO analyses.
Band 6 is entirely new, and the $M1/E2$ nature of the
in-band 363-, 367-, 442-,~423-, 468-, and 491-keV transitions was deduced
from the normalized ratios of $\gamma$-ray intensities in the detectors at forward angles
to the intensities in the detectors at angles centered around $90^{\circ}$.
This band forms a chiral pair with Band 5 and decays into it
via a number of strong $\Delta I $ = 1 and some weak $\Delta I$ = 2
transitions. The $M1/E2$ character of the 711- and 779-keV interconnecting
transitions was again deduced from an angular-distribution analysis wherein the
$A_{2}/A_{0}$ ratios were determined to be  $-(0.622\pm 0.013)$
and $-(0.7312\pm 0.032)$, respectively. In addition, the DCO ratios
for the 1074- and 1146-keV transitions were found to be consistent with a
stretched-$E2$ character~\cite{daniel.thesis}. These results establish that Band 6 has the same
(negative) parity as Band 5. Combined with the presence of strong interconnecting
transitions, this observation also points to both bands having the same intrinsic configuration.

\begin{figure}[h!]
  \begin{center}
    \includegraphics[scale=0.38]{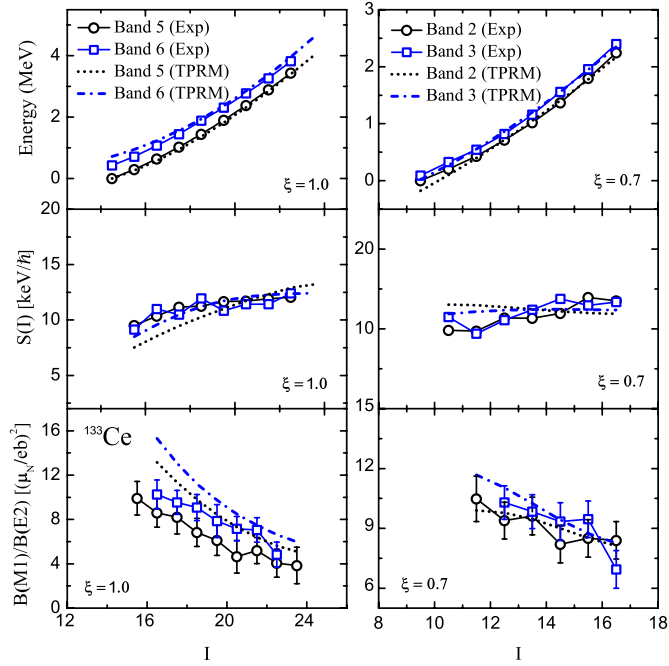}
\vspace*{-0.3 cm}
   \caption{\label{fig:epsart4}(color online) Experimental excitation energies, $S(I)$ parameters, and 
          $B(M1)/B(E2)$ ratios for the
          negative-parity chiral doublet (left panels) and positive-parity chiral doublet
          (right panels) in $^{133}$Ce. Also shown are results of TPRM calculations
          with the indicated attenuation factors $\xi$ (see text). }
\label{bm1-be2}
\end{center}
\end{figure}

The standard fingerprints for chiral bands outlined in Ref. ~\cite{starosta.aip}, i.e., close
excitation energies, a constant staggering parameter, $S(I)$, and
a similar behavior of the $B(M1)/B(E2)$ ratios, are evident for both chiral pairs in Fig.~\ref{bm1-be2}. 
Thus, there are {\em two} pairs of composite chiral partners, Bands 2--3 and Bands 5--6, in the same nucleus.
This is the first instance of firm observation of more than one set of chiral-partner bands in any nucleus. 
It should be pointed out that important confirmation of the chiral nature is provided by the measurement of transition
probabilities in the two bands, as was clearly established in case of the chiral bands in $^{135}$Nd \cite{MukhopadhyayPRL.99.172501}.
Although lifetime measurements were not feasible for bands reported here,
the identical configuration of the chiral band-pair in $^{135}$Nd and Bands 5--6
in $^{133}$Ce (see discussion below) provides further credence to these bands being chiral as well.

\ \\

\begin{table*}
\begin{center}
\caption{The unpaired-nucleon
configurations and parities for states  A, a, and b, calculated with RMF. 
The deformation parameters $\beta$
and $\gamma$, and the excitation energies $E_{\rm x}$ (in MeV) are provided, along with
the experimental bandhead energies of Bands 2 and 5, based on the configurations a and b
(see Fig.~\ref{fig4}). The spin
and parity $I^\pi$ for bandheads of the two bands are also listed.}\label{table1} 
\ \\
\begin{tabular}{ccccccccccccccccccc}
  \hline
  \hline
   & & & &\multicolumn{2}{c}{RMF calculations} &  & & \multicolumn{3}{c}{Experiment} \\
  \cline{5-6} \cline{8-11}
State  & Unpaired nucleons & Parity & &$(\beta,\gamma)$  &  $E_{\rm x}$  & & & Band  & $I^\pi$  &  $E_{\rm x}$\\
\hline
A &  $\nu(1h_{11/2})^{-1}$ & $-$ & &$(0.20, 11.1^\circ)$ & $0.00$  & & & & \\
a &  $\pi(1g_{7/2})^{-1}(1h_{11/2})^1\otimes \nu(1h_{11/2})^{-1}$ & $+$ & &$(0.22, 17.5^\circ)$ & $1.71$  & & & Band 2 & $19/2^+$ & $2.378$\\
b & $\pi(1h_{11/2})^2\otimes \nu (1h_{11/2})^{-1}$ & $-$ & &$(0.23, 15.2^\circ)$ & $3.78$  & & & Band 5 & $29/2^-$ & $3.734$\\
  \hline
\end{tabular}
\end{center}
\vspace*{-0.5 cm}
\end{table*}

\begin{figure}[h!]
  \centering
    \includegraphics[scale=0.36]{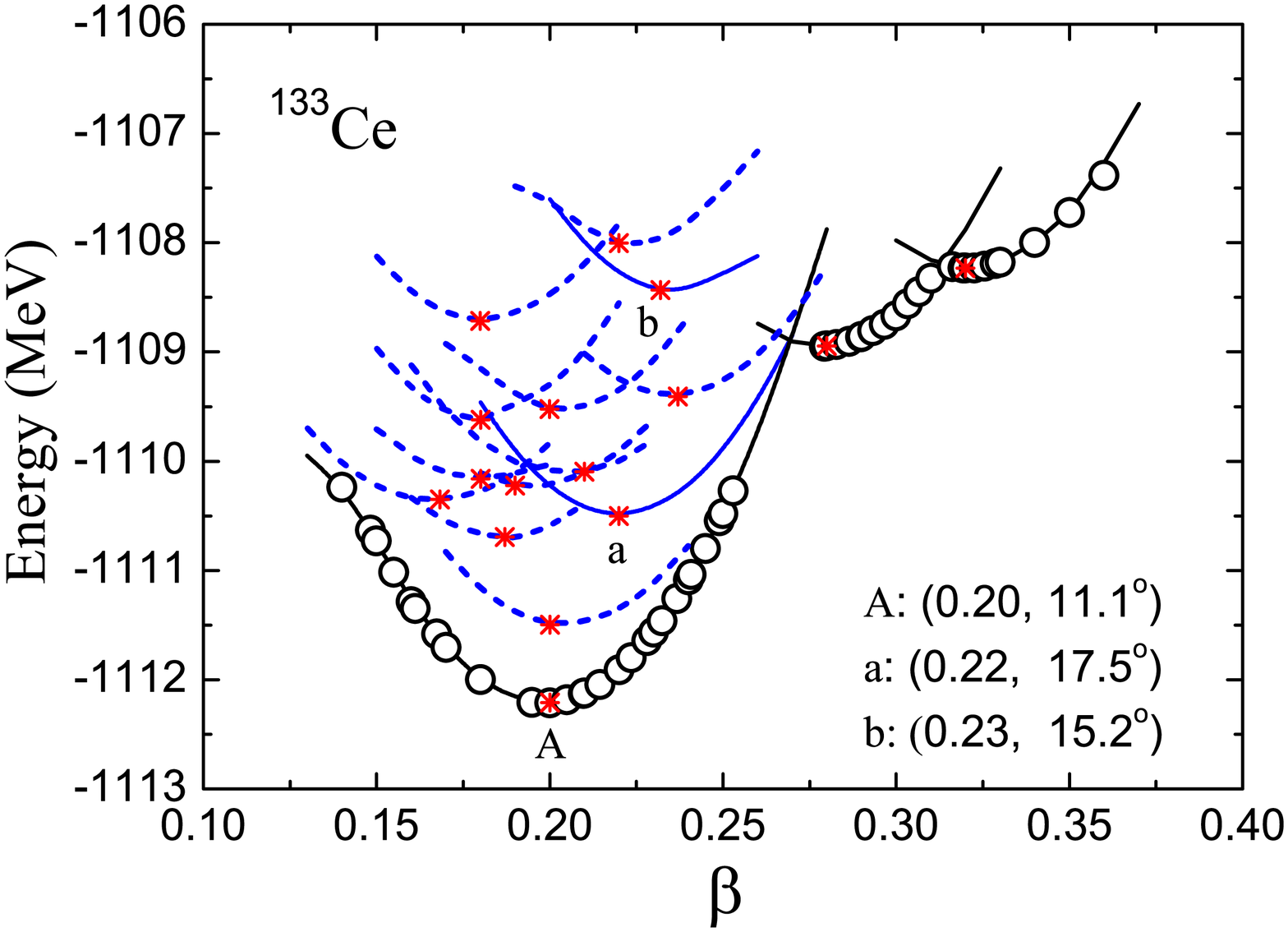}
 \vspace*{-0.3 cm}
   \caption{(Color online) Potential-energy curves in adiabatic (open circles) and various configuration-fixed 
           $\beta$-constrained  (solid and dashed curves) RMF calculations with the parameters set
           PK1~\cite{longPRC.69.034319} for $^{133}\rm Ce$. The energy minima are
           represented as stars. The letters A, a, and b denote the calculated ground
           state, Band 2, and Band 5, respectively.}
    \label{fig4}
\end{figure}

Calculations based on a combination of the constrained triaxial RMF theory~\cite{mengPRC.73.037303,pengPRC.77.024309,J.Li2011PRC} and the TPRM~\cite{S.Q.Zhang2007PRC,B.Qi2009PLB} have been performed to investigate the  nature of the observed pairs of coupled bands
and the presence of M$\chi$D phenomenon in $^{133}$Ce. The potential-energy surface in the $\beta$-$\gamma$ plane
for $^{133}\rm Ce$ obtained from the RMF calculations with the parameter set PK1~\cite{longPRC.69.034319} indicates that the calculated ground state of $^{133}\rm Ce$
is triaxial $(\beta_{2}=0.20, \gamma=11^\circ)$ and
is soft with respect to the $\gamma$ degree of freedom.
By minimizing the energy with respect to the deformation $\gamma$, both the adiabatic
and configuration-fixed $\beta$-constrained RMF calculations for $^{133}\rm Ce$
have been performed for various low-lying particle-hole excitations;
the results are provided in Fig.~\ref{fig4}. 
The configuration, parity, deformation, and energy information for the calculated ground state (labeled A in Fig.~\ref{fig4}), as well as for states ``a'' and ``b'' (regarded as bandheads of Bands 2 and 5, respectively), are presented in Table~ \ref{table1} and compared with the
experimental bandhead energies, and spins and parities $I^\pi$ of the respective bands.
The observed bandhead energies are reproduced reasonably well.

In order to examine the formation of chiral structures based on
configurations a and b, the deformation parameters $\beta$ and
$\gamma$ obtained from the RMF calculations were
used as inputs to the TPRM calculations~\cite{B.Qi2009PLB,B.Qi2011PRC}.
The moment of inertia of the rotor was adjusted to the experimental
energies and the calculation of the electromagnetic transition
probabilities followed the process described in Refs.~\cite{B.Qi2009PLB,B.Qi2011PRC}.

The $\pi(1h_{11/2}){}^{2}\otimes \nu(1h_{11/2}){}^{-1}$ quasiparticle
configuration assigned to 
Band 5 is due to the rotational alignment of
a pair of $h_{11/2}$ protons coupled to an $h_{11/2}$ neutron after band crossing. A similar observation
was made in Ref.~\cite{Ma-PhysRevC.36.2322}, but the configuration
of the band was suggested to be associated with a stable prolate axial
($\gamma$=0$^{\circ}$) shape. However, configuration assignments there
were obtained with the assumption that the axis of rotation must coincide
with one of the principal axes of the density distribution. 
In the present RMF calculations, the prolate minimum
of the previous cranking calculations becomes a saddle point, and the actual
minimum corresponds to a triaxial deformation capable of forming a chiral band.

TPRM calculations have been performed for Band 5 and its proposed chiral partner,
Band 6, based on a $\pi(1h_{11/2}){}^{2}\otimes \nu(1h_{11/2}){}^{-1}$
configuration and the results are presented in the left section of Fig.~\ref{bm1-be2}.
The theoretical results show an impressive agreement with the data. 
The energy separation between the two bands is about 400 keV
at $I=\frac{29}{2}$, and decreases rather smoothly with
increasing spin, becoming as small as 200 keV at
$I=\frac{41}{2}$. Indeed, a similar band structure built on the
$\pi(1h_{11/2}){}^{2}\otimes \nu(1h_{11/2}){}^{-1}$ configuration
was previously observed  in $^{135}$Nd and
interpreted as a chiral composite structure~\cite{PhysRevLett.91.132501,MukhopadhyayPRL.99.172501}. The nucleus $^{133}$Ce contains
two fewer protons and the energy variation between Bands 5 and 6 in
Fig.~\ref{fig:epsart4} similarly suggests a form of chiral vibration,
a tunneling between the left- and right-handed configurations,
such that Band 5 is associated with the zero- and Band 6 with the
one-phonon state. 
Furthermore, the calculated staggering parameter $S(I)$ is seen to vary
smoothly with spin (Fig.~\ref{bm1-be2}). This is expected since the Coriolis interaction
is substantially reduced for a three-dimensional coupling of angular momentum vectors in a chiral geometry. A comparison of the calculated electromagnetic transition probability ratios with the data is presented in the lower panels of Fig.~\ref{bm1-be2}.
In the TPRM calculations, no obvious odd-even staggering of the
$B(M1)/B(E2)$ values is indicated, although a small
effect is apparent experimentally. This lack of staggering 
further supports the interpretation in terms of a chiral vibration between
Bands 5 and 6~\cite{Qi-PRC.79.041302}, which may be
due to the softness of the triaxial shape~\cite{joshi.PRL.98.102501}.

The positive-parity chiral doublet bands built on the $19/2^{+}$
state (Bands 2 and 3) are associated with the three-quasiparticle 
configuration $\pi(1g_{7/2}){}^{-1}(1h_{11/2}){}^{1}
\otimes \nu(1h_{11/2}){}^{-1}$. 
The energy separation between Bands 2 and 3 was found to be nearly constant at $\sim$100 keV, which,
combined with the spin-independent $S(I)$ parameter, leads to the interpretation of these  bands being chiral partners as well.
It should be noted that the two M$\chi$D configurations in $^{133}$Ce are analogous to the 
$\pi(1g_{9/2}){}^{-1}\otimes \nu(1h_{11/2}){}^{1} (1g_{7/2}){}^{1}$ and $\pi(1g_{9/2}){}^{-1} \otimes \nu(1h_{11/2}){}^{2}$ configurations proposed previously for $^{105}$Rh~\cite{PhysRevC.69.024317,Timar2004178}. 

The calculated excitation energies and the electromagnetic
transition probability ratios $B(M1)/B(E2)$ for Bands 2 and 3 
are compared with the experimental values in the right
panels of Fig.~\ref{bm1-be2}. It can be seen that the theoretical results are
able to reproduce the data well: In the chiral region,
the energy separation between Bands 2 and 3 is nearly constant  at
$\sim$100 keV, and the similar behavior of $B(M1)/B(E2)$ ratios is also clearly evident. 
Similar to Bands 5 and 6, there is no perceptible staggering in the calculated $B(M1)/B(E2)$ ratios, but  such an effect is apparent experimentally. The B(M1)/B(E2) ratio depends sensitively on the details of the transition from the vibrational to the tunneling regime \cite{B.Qi2009PLB}, which may account for the deviations of the PRM calculation from experiment.

When comparing the experimental energies of Bands 2 and 3 with the PRM calculations,
a Coriolis attenuation factor, $\xi$=0.7, was employed.
The incorporation of such an attenuation factor is not uncommon in the description
of the properties of low-lying bands in deformed odd-$A$ nuclei with PRM
\cite{ring2} and might be a consequence of the fact that the configuration for these bands contains a low-$j$ orbital $\pi(1g_{7/2})$ which has large admixtures with other low-$j$ orbitals. The configuration for Bands 5 and 6, on the other hand, contains only the high-$j$ orbital $\pi (1h_{11/2})$, where the mixing is relatively small, thus alleviating the need for Coriolis attenuation.

IIn summary, two distinct chiral doublet bands based on the three-quasiparticle configurations
$\pi((1g_{7/2}){}^{-1} (1h_{11/2}){}^{1})\otimes \nu(1h_{11/2}){}^{-1}$
and $\pi(1h_{11/2}){}^{2}\otimes \nu(1h_{11/2}){}^{-1}$, respectively, have been
observed for the first time in $^{133}$Ce, and are interpreted in
the context of the M$\chi$D phenomenon.
Calculations based on a novel combination of the RMF and TPRM reproduce
the experimental results well, and confirm the manifestation
of triaxial shape coexistence in this nucleus.

We thank C. R. Hoffman, C. Nair, and I. Stefanescu for their help with these measurements.
UG acknowledges the Peking University Global Visiting Professors Program for a Fellowship during his sojourn at Peking University. This work has been supported in part by the U. S. National Science
Foundation (Grant Nos. PHY07-58100, PHY-0822648, and PHY-1068192); the U. S. Department of Energy, Office of Nuclear Physics, under Grant Nos. DE-FG02-95ER40939 (UND) and DE-FG02-94ER40834 (UM), and Contract 
No. DE-AC02-06CH11357 (ANL); the Major State 973 Program of China (Grant No. 2013CB834400);
the National Natural Science Foundation of China (Grant Nos. 10975007, 10975008, 11105005, and 
11175002); the Research Fund for the Doctoral Program of Higher Education, China (Grant No. 20110001110087); and the China Postdoctoral Science Foundation (Grant No. 2012M520101). 
%\newpage

\bibliographystyle{apsrev}

\bibliography{ND_MXD}

\end{document}